# Teaching electromagnetism through demonstration of a practical application involving learning content from multiple disciplines

Jiro Kitagawa

Department of Electrical Engineering, Faculty of Engineering, Fukuoka Institute of Technology, 3-30-1 Wajiro-higashi, Higashi-ku, Fukuoka 811-0295, Japan

j-kitagawa@fit.ac.jp

**Abstract**

Teaching electromagnetism by demonstrating a practical application associated with learning content is an important teaching technique. Demonstrations are often performed through video-assisted procedures, and are usually limited to a one-to-one correspondence between the application and the learning content. However, understanding of practical applications frequently requires understanding of learning content beyond the ones studied in electromagnetism. I designed a process to teach the working principles of cryogen-free superconducting magnets. Understanding of this practical application requires students to have knowledge of learning content from the disciplines of electromagnetism and solid state physics. The teaching process involved the use of homemade videos demonstrating magnet operation, which was combined with a flipped classroom and an active learning approach.

## 1. Introduction

Information, communication, and computational technologies are applied widely in education[1]. Some technologies have been utilized to deliver content, making use of the internet, simulations, games, and personal response systems, such as clickers [2,3,4,5,6]. Demonstration videos are a technology that is often used to explain to students how a fundamental natural law is related to a practical application. Demonstration videos usually show a one-to-one correspondence between the learning content and the application. However, understanding of practical applications frequently requires understanding of learning content from multiple disciplines. I have tried to make educational demonstration videos to fulfill this requirement in an undergraduate course in electrical engineering. I focused on superconducting magnets, which have been widely used in magnetic resonance imaging (MRI), linear motor cars, and other practical applications. In the superconducting magnet, a magnetic field in a coil carrying a steady current can be derived using the Biot-Savart law, which is an important element in the learning content of an electromagnetism class. Furthermore, one needs to know the physical properties of superconductors, which is usually taught in the solid state physics class. In this paper, I report on the process of teaching the working principles of cryogen-free superconducting magnets in an electromagnetism class. This process involved the use of homemade videos demonstrating magnet operation, which was combined with a flipped classroom and an active learning approach.

## 2. Flipped classroom

I have been teaching electromagnetism with the pedagogical approach of the flipped classroom[7,8], attracting much attention. Students watch prerecorded videos, which are uploaded to a commercial web system. The videos explain fundamental concepts of electromagnetism, and students watch them outside class time. A screenshot of the web system is shown in Fig. 1(a). In class, I check students' understanding of the concepts explained in the videos, via a quiz. An example of a question from the quiz is shown in Fig. 1(b). Students give their answers via clickers. Statistics of their answers are immediately displayed on the screen. After checking students' understanding of the videos, I give the students an advanced exercise, which allows them to apply the knowledge they have learned from the videos.

## 3. Superconducting magnet

Figure 2(a) is a photograph of the cryogen-free superconducting magnet (Axis, Mag 6T-52) in my laboratory. The magnet is a solenoid coil made of a superconducting niobium-titanium (Nb-Ti) wire with a diameter of 0.6 mm (see the inset of Fig. 2(a)). The schematic view of the coil is shown in Fig. 2(b). The inner ($2a_1$) and outer ($2a_2$) diameters are 90 mm and 154 mm, respectively. The length ($l$) of the coil is 103 mm. The number of coil turns ($N$) is 8,848. The generated magnetic field $B$ at the center of the coil, carrying a steady current $I$, can be calculated using the following equation:

$$B = \frac{\mu_0 NI}{2(a_2-a_1)} \ln \frac{a_2+\sqrt{a_2^2+l^2/4}}{a_1+\sqrt{a_1^2+l^2/4}} \quad (1)$$

where $\mu_0$ is the permeability of free space. Equation (1) can be derived using the Biot-Savart law. The maximum $B$ of our magnet is 6 T, generated by $I$ at 85.5 A.

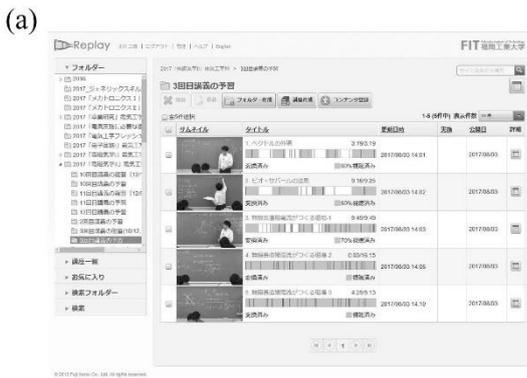

(b) Select the correct formula of Ampère's law

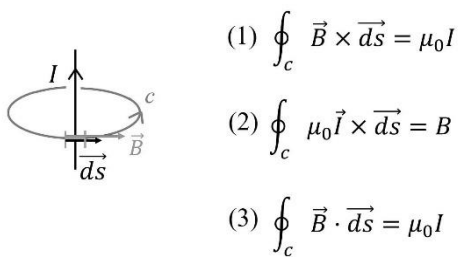

Fig. 1 (a) Screenshot of the web system where prerecorded videos are uploaded. (b) Example of a question used to check students' understanding of the concepts explained in the videos. Actual quiz is written in Japanese.

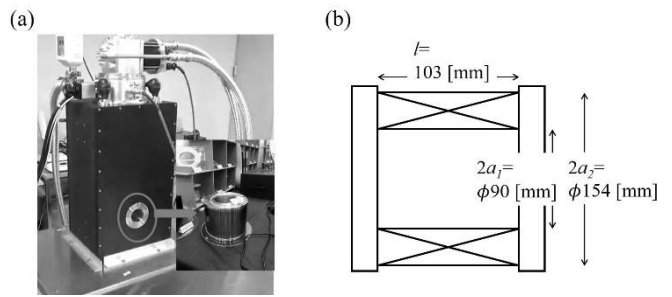

Fig. 2 (a) Photograph of cryogen free superconducting magnet. The inset is the solenoid coil in the system. (b) Schematic view of coil.

The superconducting coil is cooled down to approximately 2.5 K using a helium gas flow cryostat (see Fig. 2(a)). To understand the working principles of superconducting magnets, students need to have an understanding of the physical properties of superconductors, in addition to the Biot-Savart law. Zero resistivity and critical current density are particularly important concepts that the students need to be familiar with.

## 4. The teaching process

Prior to the class on superconducting magnets, students have learned about the Biot-Savart law, Ampère's law, and other topics in the electromagnetism class. They have also learned about the physical properties of superconductors in the solid state physics class. To teach the topic of superconducting magnets, I partially employed an active learning approach[9,10,11] in which students were given the opportunity to conduct group discussions of a given problem.

### 4.1 Introduction of superconducting magnet

As most students had no prior knowledge in the topic of superconducting magnets, I introduced the topic by presenting two practical applications of superconducting magnets: MRI and linear motor cars. To emphasize further the practical applications of superconducting magnets, I showed a demonstration video of a linear motor car. The video was readily available from the internet.

### 4.2 Limitations of the magnetic field generated by electromagnets

Before teaching the topic of superconducting magnets, I presented to the students a Weiss-type electromagnet (Tamagawa Seisakusyo, TM-YSF8615RC-083), which is in my laboratory (see Fig. 3(a)). I showed a 3-minute homemade video of the operation of the electromagnet. I explained the working principles of electromagnets, emphasizing the $B$-$H$ (where $H$ is the external field) curve of the ferromagnetic material used in the yoke (see Fig. 3(b)). Then I asked the students the following question:

Question 1: Can we generate a magnetic field that is greater than 2 T with an electromagnet?

The students discussed in groups, and then gave their answers (yes or no) via clickers. The answer is in the negative because the magnetization of ferromagnetic material becomes saturated, and $B$ approaches an

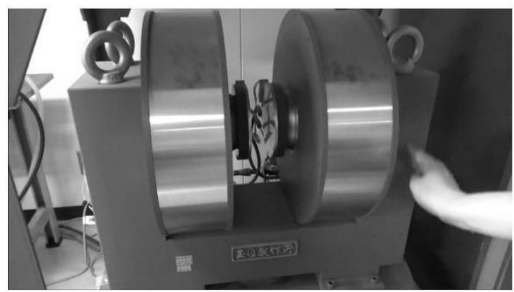

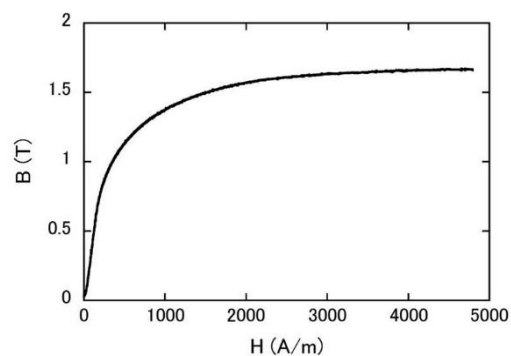

Fig. 3 (a) Photograph of electromagnet. (b) Typical $B$-$H$ curve of ferromagnetic materials.

asymptotic maximum value with increasing $H$. At this stage, students understood that the generated magnetic field has an upper limit value, which is insufficient to be used in MRI and linear motor cars.

### 4.3 Physical properties of superconductor

I explained to the students that we have stopped using ferromagnetic materials to generate magnetic fields that are greater than 2 T. I further explained that we have considered using air core coils, but coil heating is the critical issue. We need a huge current to generate a large magnetic field, but this heats up the conventional metallic wire used in air core coils. In contrast, superconductors possess the superior property of zero resistance, which is free from the problem of heating. At this stage, I reviewed the characteristic features of superconductors. This topic has already been taught in the solid state physics class. I provided explanations of zero resistivity, Meissner effect, and type I and II superconductors. I also explained the important physical parameter of critical current density $J_c$, which originates from the Lorentz force in type II superconductors. Lorentz force has already been taught in the electromagnetism class. I told the students that the superconducting Nb-Ti wire used in our magnet has a critical temperature of 9.8 K. With a magnetic field $B$ of 6 T, $J_c$ of the Nb-Ti wire is $10^5$ A/cm$^2$. Thus, with a diameter of 0.6 mm, the wire can carry a maximum current of 280 A, which is sufficient for generating a magnetic field $B$ that is greater than 2 T. Through the teaching process, I demonstrated to the students the advantages of superconducting wire.

### 4.4 Estimation of current for the generation of 6 T

I presented the design of our superconducting magnet (see Fig. 2(b)). I explained the operation of the superconducting magnet, using a prerecorded, 3-minute homemade video (Fig. 4). The video demonstrates the need of a cooled coil and the effortless generation of a magnetic field $B$ that is greater than 2 T. I showed the students the simple formula $B=\mu_0 nI$, where $n$ is the number of coil turns per unit length, which has a value of $8.6 \times 10^4$ turns/m for our magnet. I performed the calculations to estimate the value of $I$, the current carried by the wire, needed to generate a magnetic field $B$ of 6 T. The calculations resulted in an estimated current of 56 A, which is smaller than the value of 89 A noted in the specification sheet of the magnet. Then I asked the students the next question:

Question 2: What is at the origin of the difference between the current obtained from $B=\mu_0 nI$ and that noted in the specification sheet?

A few groups of students concluded that it was because the length of the superconducting coil is finite, while the formula is derived for a solenoid coil with infinite length.

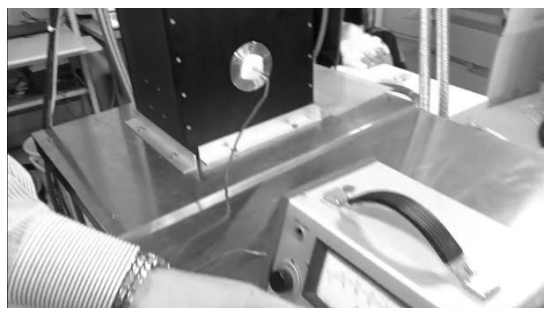

Fig. 4 Screenshot of homemade video explaining the operation of the superconducting magnet.

Finally, students in groups considered the following question:

Question 3: How can we obtain estimates of the current with higher accuracy?

The question is difficult, and after 15~20 minutes of group discussion (see Fig. 5(a)) and presentation by a student representative (see Fig. 5(b)), I presented Equation (1) to the students. It is a formula that is more precise and it can be derived using the Biot-Savart law.

(a)
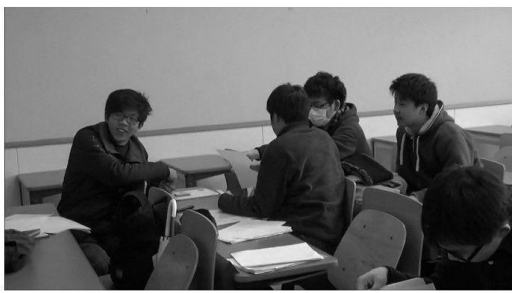

(b)
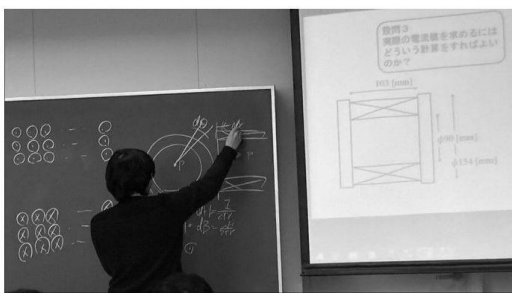

Fig. 5 (a) Photograph of group discussions around Question 3. (b) Photograph of a student presenting the results of the group discussions.

## 4. 5 Summary

I have reported on a teaching process that involved the educational demonstration of a cryogen-free superconducting magnet. To understand the working principles of superconducting magnets, students need knowledge of electromagnetism and solid state physics. The demonstration was conducted through homemade videos, combined with an active learning approach. Students conducted group discussions on the limitations of conventional electromagnets in generating large magnetic fields, and on the calculation of magnetic fields generated by superconducting magnets. Furthermore, students learned about the advantages of superconducting wire in generating the large magnetic fields needed for MRI and linear motor cars. I hope many students have become aware that, to understand practical applications of fundamental natural laws, understanding of learning content from multiple disciplines is necessary.

## 5. Acknowledgement

I acknowledge the Acceleration Program for University Education Rebuilding from the Ministry of Education, Culture, Sports, Science and Technology for the support of system FITReplay, which made the uploading of the prerecorded videos possible. I am also grateful for the financial support from the budget committee of Fukuoka Institute of Technology. This work was performed under our accepted projects "Introduction of clickers in the flipped classroom" and "Introduction of cryogen-free superconducting magnets, aiming at advanced education and research". I thank Tina Tin, PhD, from Edanz Group (www.edanzediting.com/ac) for editing a draft of this manuscript.

---